\documentclass[%
 reprint, 
 superscriptaddress,
 amsmath,amssymb,
 aps,
floatfix,
longbibliography
]{revtex4-2}
\usepackage{graphicx} 
\usepackage{amsmath}
\usepackage[colorlinks=true,citecolor=blue,linkcolor=magenta]{hyperref}
\usepackage{braket}
\usepackage{enumitem}

\allowdisplaybreaks 

\begin{document}

\title{Application-Aware Benchmarking on NISQ Hardware \protect\\ using Expectation Value Fidelities}
\author{Joseph Harris}
\email{joseph.harris@dlr.de}

\author{Peter K. Schuhmacher}

\affiliation{Department of High Performance Computing, Institute of Software Technology, German Aerospace Center (DLR),  Rathausallee 12, 53757 Sankt Augustin, Germany}
\date{March 2025}

\begin{abstract}
    We present a low-cost protocol for benchmarking applications on generic quantum hardware in the circuit model. Using families of Clifford circuits which mimic the application circuit structure, we are able to predict how measured expectation value fidelities scale with circuit depth. We consider the specific example of simulating a kicked-Ising model on superconducting hardware, showing our benchmark to be more accurate than predictions which use the gate error data obtained through randomized benchmarking. We also demonstrate how our work can be used to benchmark the performance and limitations of quantum error mitigation techniques. Our method is targeted at applications which have a natural decomposition in terms of Pauli rotations, but can be applied to any input circuit with this decomposition. 
\end{abstract}

\maketitle

\section{Introduction}

The performance of a noisy intermediate-scale quantum (NISQ) computer in implementing a given algorithm can be measured in several ways. For example, the implementation of the full unitary on hardware can be completely characterized using full quantum process tomography \cite{Chuang01111997, PhysRevLett.78.390}, although this procedure is not scalable with an increasing number of qubits. More commonly, randomized benchmarking is used to estimate the average fidelities of individual native gates which are implemented on hardware \cite{PhysRevA.77.012307, Wallman2014, PhysRevLett.123.030503, PhysRevLett.123.060501, PRXQuantum.3.020357}. 
These fidelities can then be multiplied to estimate the fidelity of an entire circuit composed of these gates. This approach has only polynomial scaling, but may not accurately describe the error behavior of hardware and does not take into account the particular application being implemented. As such, there also exist several application-tailored benchmarks \cite{9773202, 10061574, Mills2021}. These approaches consider various figures of merit to quantify performance. Some treat generic circuits while others consider only specific applications or subroutines, relying on classical knowledge of the input problem.

Our aim is to introduce a benchmarking method which is as application and hardware-agnostic as possible, whilst still being able to accurately capture the effects of a device's complex error behavior on the input circuit when compared to existing benchmarking approaches. In this work, we consider the implementation of a generic circuit which can be decomposed straightforwardly in terms of Pauli rotations, followed by an expectation value measurement of a Pauli observable. In particular, for a given quantum application circuit $\mathcal{U}$ and Pauli observable $O$ we define the expectation value fidelity
\begin{equation}
    F_{\mathcal{U}, O} = \frac{\braket{O}_\mathcal{U}^\text{noisy}}{\braket{O}_\mathcal{U}^\text{ideal}}
\end{equation}
as our figure of merit. 
This quantity is straightforward to interpret: values close to one indicate good performance, and we expect $F_{\mathcal{U}, O}$ to decay exponentially with circuit depth. We then show how to construct, given $\mathcal{U}$, a family of Clifford benchmarking circuits $B_\mathcal{U}$ such that 
\begin{equation}
    F_{\mathcal{B}, O} \approx F_{\mathcal{U}, O}
    \label{eq:fidelity-estimate}
\end{equation}
for all $\mathcal{B} \in B_\mathcal{U}$, under specific assumptions which we discuss in the next section. 
By the Gottesman–Knill theorem \cite{osti_319738}, our method is scalable to large numbers of qubits $n$ since our Clifford benchmarking circuits may be classically simulated in $\text{poly}(n)$ time.

As we will demonstrate, this construction can result in more accurate fidelity estimates when compared to predictions using gate fidelity data obtained through randomized benchmarking. By mimicking the application circuit structure, our benchmarking circuits capture more complex error behavior than is predicted through randomized benchmarking such as non-local crosstalk \cite{Sarovar2020detectingcrosstalk}. 

To demonstrate our method, we consider the quantum simulation of a two-dimensional kicked Ising chain followed by a measurement of the magnetization ($Z$) of a single site, using a superconducting processor. This application was recently used by IBM and others \cite{Kim2023, PRXQuantum.5.010308, doi:10.1126/sciadv.adk4321, KECHEDZHI2024431, anand2023classical, shao2023simulating, liao2023simulation, Rudolph:2023ayj} in the context of demonstrating the \textit{utility} of quantum hardware. Hence, we believe benchmarking the performance of this application to be of genuine interest; additionally, the application has a straightforward compilation in terms of Pauli rotations. 

This work is structured as follows. In Section \ref{sec:generating-benchmarking-circuits}, we describe how to generate a family of Clifford benchmarking circuits $B_\mathcal{U}$ from a single application circuit $\mathcal{U}$ and how to use these circuits to estimate the expectation value fidelity $F_{\mathcal{U}, O}$. In Section \ref{sec:benchmarking-ki-on-superconducting} we numerically demonstrate our protocol, benchmarking the performance of kicked-Ising simulations on superconducting hardware as well as the performance of an error mitigation protocol. In the final sections we discuss the outlook of our work. Alongside this paper, we provide a GitHub repository containing all code and data produced for this work \cite{Github}.

\section{Generating benchmarking circuits}
\label{sec:generating-benchmarking-circuits}

We begin with some definitions which will be useful to us later. A \textit{Pauli string} $P$ of length $n$ is a tensor product $P = Q_1 \otimes \cdots \otimes Q_n$ where each $Q_i \in \{I,X,Y,Z\}$ is a standard Pauli matrix. We define the \textit{$n$-qubit Pauli rotation gates} $R_P(\theta) = e^{-i\theta P/2}$, where $P$ is a Pauli string of length $n$. We also define $T(P) = \{i \ |  Q_i \neq I\}$ to be the set of qubits on which $P$ acts non-trivially, e.g. $T(IIXIZ) = \{3,5\}$, and the \textit{weight} $w(P) = |T(P)|$ to be the number of such qubits. 

For $P \in \{X,Y,Z\}$ we also define $\ket{P^{\pm 1}}$ to be the eigenstate of $P$ with eigenvalue $\pm1$, so that for example $\ket{Z^{+1}} = \ket{0}$ and $\ket{X^{-1}} = \ket{-}$. 

The set of all single and two-qubit Pauli rotations is universal for quantum computing; hence we consider for simplicity our $n$-qubit input application circuit $\mathcal{U}$ to be decomposed in terms of $M$ such rotations
\begin{equation}
    \mathcal{U} = R_{P_M}(\theta_M) \cdots R_{P_1}(\theta_1),
\end{equation}
where each $P_i$ has length $n$ and weight 1 or 2. Our work can be straightforwardly extended to multi-qubit gates if needed.

Suppose that $\mathcal{U}$ is an $n$-qubit circuit with input state $\ket{0}^{\otimes n}$. To generate a single Clifford benchmarking circuit $\mathcal{B} \in B_\mathcal{U}$, we first start with the ansatz 
\begin{equation}
    R_{\tilde{P}_M}(\tilde{\theta}_M) \cdots R_{\tilde{P}_1}(\tilde{\theta}_1),
\end{equation}
and pick the Paulis $\tilde{P}_i$ and angles $\tilde{\theta}_i$ via the following algorithm:
\begin{enumerate}
    \item Let $\ket{\psi_{q}}$ denote the state of qubit $q$ and set $\ket{\psi_{q}} = \ket{Z^{+}}$ for all $q = 1, \dots, n$. We will update these as we move through the circuit to keep track of the overall circuit state, which remains a product state throughout. 
    \item For $i = 1, \dots, M$:
    \begin{enumerate}
        \item If $w(P_i) = 1$, pick uniformly at random $\tilde{\theta}_i \in \{0,\pi/2, \pi, 3\pi/2\}$ and $\tilde{P}_i \in \{X,Y,Z\}$. Let $T(P_1) = \{a\}$ and update $\ket{\psi_{a}} \mapsto  R_{\tilde{P}_i}(\tilde{\theta}_i) \ket{\psi_{a}}$;  this will produce another Pauli eigenstate.
        \item If $w(P_i) = 2$, pick uniformly at random $\tilde{\theta}_i \in \{0,\pi\}$.
        \begin{enumerate}
            \item If $\tilde{\theta} = 0$, pick $Q_1, Q_2 \in \{X,Y,Z\}$ each uniformly at random and set $\tilde{P}_i = Q_1 \otimes Q_2$.
            \item If $\tilde{\theta} = \pi$, let $T(P_i) = \{a,b\}$, where the qubit labels $a$ and $b$ are chosen uniformly at random. Write the current state of qubit $a$ as $\ket{\psi_{a}} = \ket{Q_a^{c_a}}$ where $Q_a \in \{X,Y,Z\}$ and $c_a = \pm 1$, and likewise for qubit $b$. Set $\tilde{P}_a = Q_a$ and pick $\tilde{P}_b \in \{X,Y,Z\}$ uniformly at random. If $\tilde{P}_b \neq Q_b$, then update $\ket{\psi_{b}} \mapsto \ket{Q_b^{-c_b}}$.
        \end{enumerate}
    \end{enumerate}
\end{enumerate}

The output state of this sequence of rotations is a known product state of Pauli eigenstates, $\otimes_{q}\ket{\psi_q}$. We can hence permute this state to be a +1-eigenstate of the Pauli observable $O = \otimes_{q=1}^n O_q$ by applying a single layer of single-qubit Pauli rotation gates:
\begin{equation}
    \prod_{q \in T(O)} R_{S_q}(\phi_q)
    \label{eq:correction-layer}
\end{equation}
where $S_q \in \{X,Y,Z\}$ and $\phi_q \in \{0, \pi/2, \pi,  3\pi/2\}$ are chosen such that $R_{S_q}(\phi_q) \ket{\psi_q} = \ket{O_q^{+1}}$, the +1 eigenstate of $O_q$. The full benchmarking circuit is then
\begin{equation}
    \mathcal{B} = \left(\prod_{q \in T(O)} R_{S_q}(\phi_q)\right) R_{\tilde{P}_M}(\tilde{\theta}_M) \cdots R_{\tilde{P}_1}(\tilde{\theta}_1).
\end{equation}

This procedure results in a Clifford circuit $\mathcal{B}$ which, compared to our original application circuit $\mathcal{U}$, has the same number of Pauli rotation gates, acting in the same places on the same qubits, plus an additional correction layer of at most $n$ single-qubit Pauli rotations at the end of the circuit. Moreover, $\mathcal{B}$ has $\braket{O}_\mathcal{B}^\text{ideal} = 1$, and hence our estimate (\ref{eq:fidelity-estimate}) 
for $F_{\mathcal{U},O}$ simplifies to
\begin{equation}
    \braket{O}_\mathcal{B}^\text{noisy} \approx F_{\mathcal{U},O} \ .
    \label{eq:fidelity-estimate-simplified}
\end{equation}
This allows us to estimate the expectation value fidelity $F_{\mathcal{U},O}$ of our application circuit by simply taking noisy measurements of $O$ with respect to our benchmarking circuits. In general, we generate many such benchmarking circuits and take an average of these measurements. This allows us to quantify not just an estimate for $\braket{O}_\mathcal{U}^\text{ideal}$ but also the variance in such estimates that we would expect across circuit runs. 

For the estimate (\ref{eq:fidelity-estimate-simplified}) to hold, we must assume that the gates $R_{P_i}(\theta_i)$ and $R_{\tilde{P}_i}(\tilde{\theta}_i)$ are subjected to very similar errors when compiled and and implemented on the target hardware. This assumption holds, for example, for any hardware platform in which
\begin{itemize}
    \item the native single-qubit [two-qubit] gates have very similar error rates;
    \item any single-qubit [two-qubit] Pauli rotation may be compiled using the same number of single and two-qubit gates in the same order.
\end{itemize}
In Section \ref{sec:rigid-compilation}, we show how these constrains can be satisfied by introducing rigid compilation rules for compiling Pauli rotations for IBM superconducting hardware. However, we particularly note that our method does not rely on a specific noise model of the quantum hardware.

\section{Benchmarking kicked-Ising circuits on superconducting hardware}
\label{sec:benchmarking-ki-on-superconducting}

\subsection{Problem statement}

\begin{figure}
    \centering
    \hspace*{-0.5em}
    \includegraphics[scale=0.55]{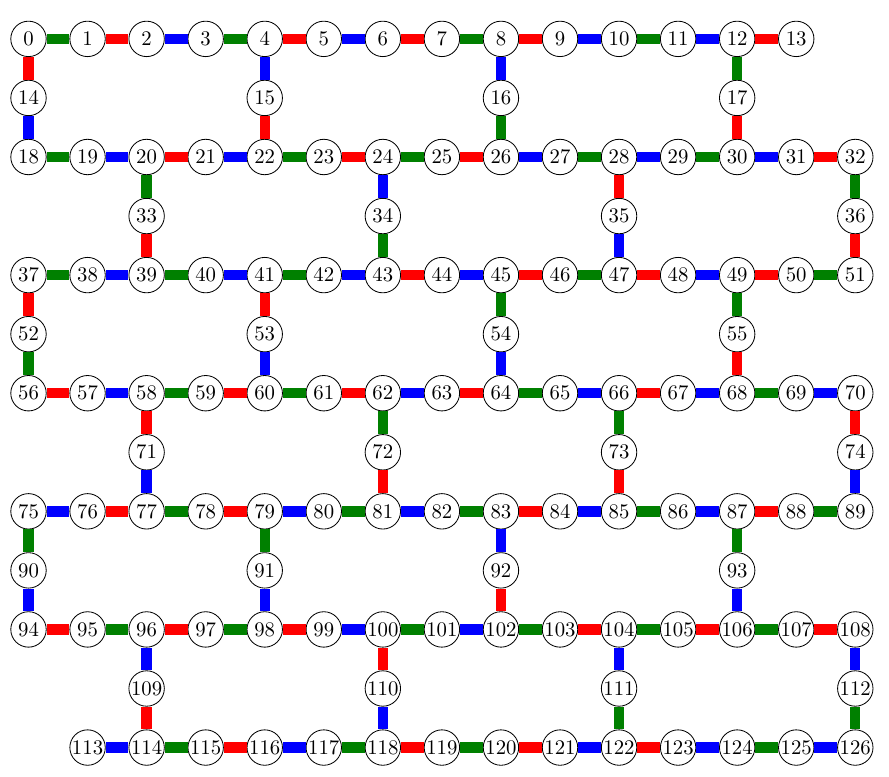}
    \caption{Qubit layout of the 127-qubit \texttt{ibm\textunderscore brisbane} device used in this work . The colors of the edges indicate the three layers of simultaneous two-qubit gates needed to implement all nearest-neighbor interactions. We apply these layers in the order \textit{red}, \textit{blue}, \textit{green}.}
    \label{fig:qubit-layout}
\end{figure}

\begin{figure*}[t!]
    \centering
    \hspace*{-0.5em}
    \includegraphics[scale=0.8]{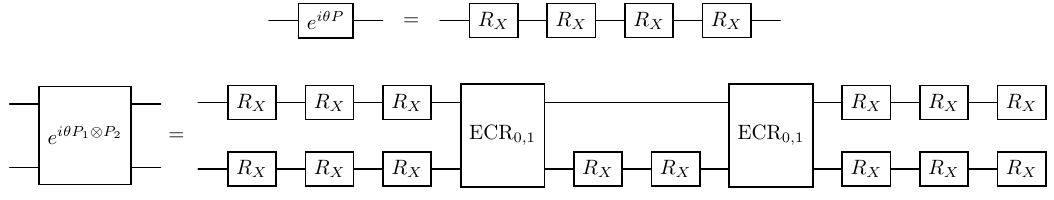}
    \caption{We compile all single-qubit Pauli rotations for the \texttt{ibm\textunderscore brisbane} device using precisely four $R_X$ gates, where $R_X \in \{\sqrt{X}, X\}$, and an arbitrary number of error-free $R_Z(\theta)$ rotations, which we emit from the picture. Similarly, all two-qubit Pauli rotations are compiled using fourteen $R_X$ gates, two $ECR$ gates and an arbitrary number of $R_Z(\theta)$ rotations.}
    \label{fig:rigid-compilation}
\end{figure*}

\begin{figure*}[t!]
    \centering
    \hspace*{-0.5em}
    \includegraphics[scale=0.7]{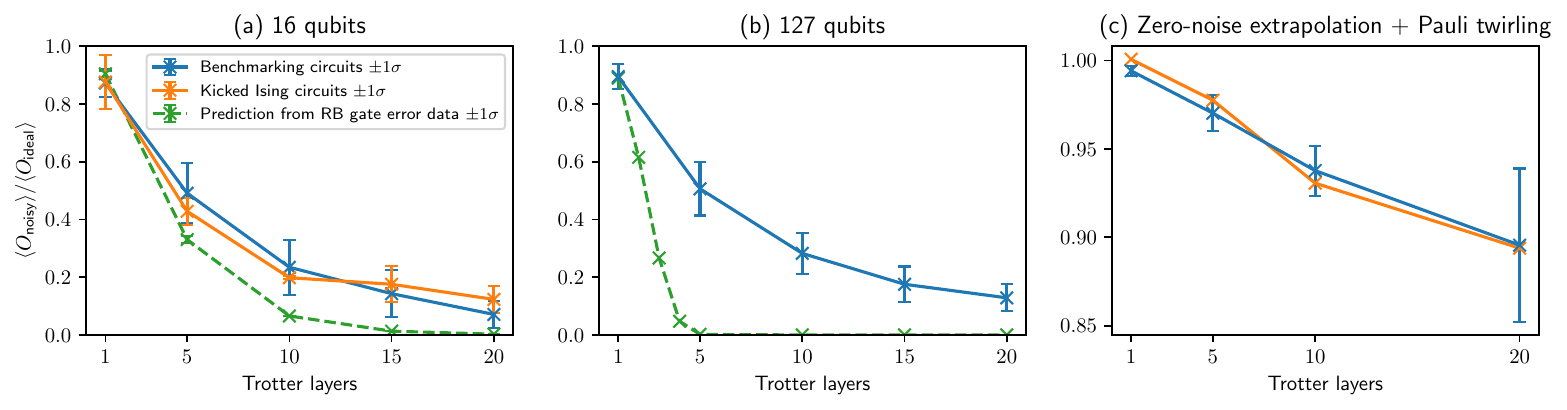}
    \caption{
    Presentation of our numerical benchmarking results. (a) and (b) were run on the \texttt{ibm\textunderscore brisbane} hardware using circuits with 16 and 127 qubits respectively. In these cases we also overlay the prediction using IBM's published gate error data obtained using randomized benchmarking. (a) was produced using 30 benchmarking circuits and 3 application circuits per data point, and shows that our benchmark is able to accurately capture the fidelity decay behavior of the application circuits. The deviation from the RB prediction also shows that the device error is not simply depolarizing. (b) uses 10 benchmarking circuits per data point and demonstrates the scalability of our method. (c) uses 10-qubit classical noisy simulation to demonstrate how our work can also be used to benchmark error mitigation methods. Here, we benchmark the efficacy of using zero-noise extrapolation and Pauli twirling for noise reduction. We use 10 benchmarking circuits and one application circuit per data point. Each circuit is run at 3 noise factors (1,3,5) with 25 Pauli twirling repetitions per noise factor. 
     }
    \label{fig:numerical-results}
\end{figure*}

To demonstrate our benchmarking protocol we consider the particular application of simulating the transverse-field (`kicked') Ising model with Hamiltonian
\begin{equation}
    H = -J \sum_{\braket{i,j}}\sigma_z^i\sigma_z^{j} + h \sum_{i}\sigma_x^i,
    \label{eq:ki-hamiltonian}
\end{equation}
where $J$ is the nearest-neighbour coupling strength and $h$ is the strength of a transverse field applied globally. The first sum runs over all pairs of nearest-neighbour qubits $(i,j)$ and the second runs over all qubits in the hardware layout. For hardware, we use the \texttt{ibm\textunderscore brisbane} superconducting processor, which has a two-dimensional heavy-hexagon topology as illustrated in Figure \ref{fig:qubit-layout}. We pick as our observable $O = Z_{62}$, such that measuring $O$ corresponds to measuring the magnetization of a single site in the center of the topology.

In the circuit model, the time evolution $e^{-iHt}$ of $H$ can be simulated using a Trotter decomposition with $T$ layers of single and two-qubit gates,
\begin{equation}
    e^{-iHt} = \left[ L_1 L_2 \right]^T + \mathcal{O}\left(\frac{t^2}{T}\right)
\end{equation}
where
\begin{equation}
    L_1 = \prod_{\braket{i,j}} R_{Z_i Z_j}\left( \frac{Jt}{2T}\right), \quad  L_2 = \prod_{i} R_{X_i} \left(-\frac{ht}{2T} \right).
\end{equation}
Hence, we take our ideal application circuit to be $\mathcal{U} = [L_1 L_2]^T$, where we vary the number of Trotter layers $T$. For simplicity we set the angle parameters such that $\frac{Jt}{2T} = -\frac{ht}{2T} = 0.01$ in each case, since this results in comparable values of $\braket{O}^\text{exact}$ as $T$ varies between 1 and 20. Each two-qubit gate layer $L_1$ is separated into three layers of simultaneously implementable gates as illustrated in Figure \ref{fig:qubit-layout}.

\subsection{Rigid compilation of application and benchmarking circuits}
\label{sec:rigid-compilation}

The hardware we consider has basis gate set $\{X, \sqrt{X}, R_Z(\theta), ECR\}$, where $\theta$ is a continuous parameter and $ECR = \frac{1}{\sqrt2}(IX- XY)$. The publicly available error rates of these individual gates are regularly calculated for each qubit using randomized benchmarking and are accessible via the IBM Quantum platform \cite{IBMQuantum}. Most notably, the $R_z$ gates may be regarded as error-free, since they are implemented `virtually' by adjusting the relative phase between the projection of the qubit vector on the Bloch sphere and the microwave control tone \cite{PhysRevA.96.022330}. The $X$ and $\sqrt{X}$ gates have the same reported error rate per qubit, and the $ECR$ gate error is typically between one and two orders of magnitude larger. 

To ensure that the error behavior between each Pauli rotation $R_{P_i}(\theta)$ and $R_{\tilde{P}_i}(\tilde{\theta}_i)$ is as similar as possible, we enforce that all Pauli rotations are compiled using the same number of $\{\sqrt{X},X\}$ and $ECR$ gates. For generic single-qubit Pauli rotations this can be achieved using four $\{\sqrt{X},X\}$  gates; for two-qubit Pauli rotations we require fourteen $\{\sqrt{X},X\}$  gates and two $ECR$ gates (plus, in either case, an arbitrary number of error-free $R_Z$ gates). We present these compilations in Figure \ref{fig:rigid-compilation}.

By comparison to the most optimal compilation, this approach increases the circuit depth (ignoring $R_Z$ gates) by a constant factor, but notably does not increase the number of native two-qubit $ECR$ gates. 

We note here that, in situations where the basis gate set consists entirely of gates which can be made Clifford, one can more efficiently generate Clifford circuits by first compiling the application circuit in terms of the basis gate set and then `Cliffordizing' the circuit. In our case, this corresponds to rounding the rotation angle of each $R_Z$ gate to the nearest multiple of $\pi/2$, since the $\sqrt{X},X$ and $ECR$ gates are already Clifford; this approach was recently introduced into Qiskit as a debugging tool \cite{IBMQuantumNeat}. However, we found that this approach is not, in general, a reliable one since one has no control over how the output state of the circuit changes after Cliffordization; often the Cliffordized circuit has $\braket{O}^\text{ideal} = 0$ due to symmetries in the output state, which could be highly entangled. Our approach is designed to circumvent this issue by maintaining full control over the function of our benchmarking circuits.

\subsection{Numerical results}

We demonstrate our protocol numerically by benchmarking the 127-qubit \texttt{ibm\textunderscore brisbane} superconducting device for running kicked Ising simulations; our results are presented in Figure \ref{fig:numerical-results}. 

In the leftmost plot, we demonstrate that our benchmarking method is able to accurately capture the expectation value decay of the application circuits. We run 16-qubit circuits centered around qubit 62 with varying numbers of Trotter layers. For each fixed number of Trotter layers, we run 30 randomly-generated benchmarking circuits and 3 kicked-Ising circuits, in each case measuring $\braket{O} = \braket{Z_{62}}$, and plot the resulting expectation value fidelity $\braket{O}^\text{noisy}/\braket{O}^\text{ideal}$. We also overlay the predicted fidelity calculated using the published gate error rates for each qubit, which are obtained using randomized benchmarking (RB). This is calculated by regarding each native gate $G_i$ with error $p_i$ as a depolarizing channel,
\begin{equation}
    \rho \mapsto (1-p_i) G \rho G^\dagger + p_i \frac{I}{2^n},
\end{equation}
where $I$ is the $2^n \times 2^n$ identity matrix, and so the effect of $L$ such gates $G_1, \dots, G_L$ is then
\begin{equation}
    \rho \mapsto \tilde{\rho} \equiv \varphi \cdot (G_L \cdots G_1) \rho (G_1^\dagger \cdots G_L^\dagger) + (1-\varphi)\cdot \frac{I}{2^n},
\end{equation}
where $\varphi = (1-p_1)\cdots (1-p_L)$. Thus the noisy expectation value is then $\braket{O}^\text{noisy} = \text{Tr}(\tilde{\rho}O) = \varphi \braket{O}^\text{ideal}$.
The deviation from the RB data demonstrates that the device errors are not simply depolarizing. These results were obtained by using the device on three separate days, allowing us to also see how the gate error data can vary in practice. 

In the middle plot, we demonstrate the scalability of our method by running benchmarking circuits which utilize all 127 qubits on the device, using 10 benchmarking circuits per data point. In this case, we cannot classically verify the accuracy of our benchmarking without relying on heavily tailored tensor network methods \cite{PRXQuantum.5.010308, doi:10.1126/sciadv.adk4321, anand2023classical, liao2023simulation}. We see a steeper drop-off in the randomized benchmarking prediction since the lightcone of $Z_{62}$ contains many more gates with this increased number of qubits. 

In the rightmost plot, we demonstrate additionally how benchmarking circuits can be used to benchmark the efficacy of error mitigation techniques. These results are simulated classically using 10-qubit noisy simulations where each gate is simulated as a depolarizing channel using the provided device gate error rates from \texttt{ibm\textunderscore brisbane}. In this case each data point uses 10 benchmarking circuits or a single kicked-Ising circuit. For error mitigation we use Pauli twirling (PT) \cite{PhysRevA.54.3824, PhysRevLett.119.180509} and zero-noise extrapolation (ZNE) with two-qubit gate folding \cite{9259940, PhysRevA.102.012426}. For each circuit we measure $\braket{O}$ at three amplified noise levels (1,3,5) to extrapolate the zero-noise value; for each fixed noise level average over 25 random Pauli twirling instances. Due to the similar error behavior enforced by our rigid compilation, we see the performance of the error mitigation method applied to the application is accurately captured by our benchmark. More generally, we are able to conclude that ZNE + PT can improve the expectation value fidelity but does not perfectly mitigate errors; we still see a decay with circuit depth. Moreover, the error bars indicate that the error mitigation is less reliable at greater circuit depths. 

\section{Conclusion}
\label{sec:conclusion}

In this work, we have introduced a novel application-level benchmark for near-term quantum computers which is both application and hardware agnostic with low overhead, placing no reliance on any particular noise model. Our work is best suited to applications which have a natural decomposition in terms of Pauli rotation gates, although any unitary circuit can be expressed in this way. Our work may also be applied to any gate-based hardware, provided there exists a consistent way of compiling one and two-qubit Pauli rotation gates in the basis gate set. We have demonstrated our protocol on superconducting hardware, showing it to outperform predictions which use error rates obtained through randomized benchmarking. We have also shown how it can be used to benchmark the efficacy of error mitigation techniques. 

Our work may be used in any context where the expectation value fidelity provides a sensible figure of merit for the performance of quantum hardware and/or software. In particular, our method provides a useful tool in determining which applications are best suited to which hardware platforms, or which error mitigation methods are best suited to which combination of application and hardware platform.


\section{Code and data availability}

All program code, simulation and measurement results produced over the course of this research are available in an associated GitHub repository \cite{Github}. Also available there is the hardware calibration data for the \texttt{ibm\textunderscore brisbane} device available at the times that our circuits were run. \\

\section{Acknowledgments}
This research was made possible by the DLR Quantum Computing Initiative and the Federal Ministry for Economic Affairs and Climate Action \cite{ALQU}.

We thank Michael Epping, Benedikt Fauseweh and Frank Wilhelm-Mauch for fruitful discussions over the course of this work. 

We acknowledge the use of IBM Quantum services for this work. The views expressed are those of the authors, and do not reflect the official policy or position of IBM or the IBM Quantum team. All quantum circuits were simulated or submitted to hardware using Qiskit \cite{qiskit2024}.

\end{document}